\UseRawInputEncoding
\documentclass[12pt,letter]{article}
\pdfoutput=1
\usepackage{graphicx, epsfig, color,cite,inputenc}
\usepackage{amsmath}
\usepackage{amssymb}
\usepackage{float}
\usepackage{caption,subcaption,graphicx}
\usepackage{hyperref}

\def\to{\rightarrow}

\def\bi{\begin{itemize}}
\def\ei{\end{itemize}}
\def\te{\tilde e}
\def\tchi{\tilde\chi}

\def\tu{\tilde u}

\def\tb{\tilde b}

\def\tst{\tilde t}
\def\ttau{\tilde \tau}
\def\tmu{\tilde \mu}
\def\tg{\tilde g}
\def\tnu{\tilde\nu}
\def\tell{\tilde\ell}

\def\tw{\widetilde\chi^{\pm}}
\def\tz{\widetilde\chi^0}
\def\alt{\lesssim}
\def\agt{\gtrsim}
\def\be{\begin{equation}}  
\def\ee{\end{equation}}  
\def\bea{\begin{eqnarray}}  
\def\eea{\end{eqnarray}}  

\def\SUSY{\text{SUSY}}  
  
\def\PU{\text {PU}}  
\def\OU{\text {OU}}

\def\EW{\text {EW}}

\begin{document}
\begin{titlepage}
\begin{flushright}
OU-HEP-210415
\end{flushright}

\vspace{0.5cm}
\begin{center}
{\Large \bf Anomalous muon magnetic moment, supersymmetry,\\
naturalness, LHC search limits and the landscape
}\\ 
\vspace{1.2cm} \renewcommand{\thefootnote}{\fnsymbol{footnote}}
{\large Howard Baer$^1$\footnote[1]{Email: baer@ou.edu },
Vernon Barger$^2$\footnote[2]{Email: barger@pheno.wisc.edu},
Hasan Serce$^1$\footnote[3]{Email: hasbarser@gmail.com} 
}\\ 
\vspace{1.2cm} \renewcommand{\thefootnote}{\arabic{footnote}}
{\it 
$^1$Homer L. Dodge Department of Physics and Astronomy,
University of Oklahoma, Norman, OK 73019, USA \\[3pt]
}
{\it 
$^2$Department of Physics,
University of Wisconsin, Madison, WI 53706 USA \\[3pt]
}

\end{center}

\vspace{0.5cm}
\begin{abstract}
The recent measurement of the muon anomalous magnetic moment 
$a_\mu\equiv (g-2)_\mu/2$ by the Fermilab Muon $g-2$ experiment 
sharpens an earlier discrepancy between theory and the BNL E821 experiment.
We examine the predicted $\Delta a_\mu\equiv a_\mu(exp)-a_\mu(th)$ in the context
of supersymmetry with low electroweak naturalness (restricting to models which
give a plausible explanation for the magnitude of the weak scale). 
A global analysis including LHC Higgs mass and sparticle search limits 
points to interpretation within the normal scalar mass hierarchy (NSMH) 
SUSY model wherein first/second generation matter scalars are much 
lighter than third generation scalars.
We present a benchmark model for a viable NSMH point which is natural,
obeys LHC Higgs and sparticle mass constraints 
and explains the muon magnetic anomaly.
Aside from NSMH models, then we find the $(g-2)_\mu$ anomaly 
cannot be explained within the context 
of natural SUSY, where a variety of data point to decoupled 
first/second generation scalars. 
The situation is worse within the string landscape where first/second
generation matter scalars are pulled to values in the $10-50$ TeV range.
An alternative interpretation for SUSY models with decoupled scalar masses
is that perhaps the recent lattice evaluation of the hadronic vacuum 
polarization could be confirmed which leads to a 
Standard Model theory-experiment agreement in which case there is no anomaly.
\noindent
\end{abstract}
\end{titlepage}

\section{Introduction}
\label{sec:intro}

The recent measurement of the muon anomalous magnetic moment 
$a_\mu\equiv (g-2)_\mu/2$ by the Fermilab Muon $g-2$ Collaboration 
experiment E989\cite{Abi:2021gix} has confirmed previous results 
from the BNL E821 experiment\cite{Bennett:2004pv} which had found a 
$3.7\sigma$ discrepancy between the experimental observation 
and the theoretical prediction, taken from the Muon $g-2$ Theory 
Initiative\cite{Aoyama:2020ynm,Davier:2019can}. The latter relies mainly on using 
dispersive techniques applied to $e^+e^-\to hadrons$ data in the vicinity
of $\sqrt{s}\sim 1$ GeV to evaluate the hadronic vacuum polarization (HVP)
contribution to $a_\mu$ which contributes the largest uncertainty to the 
overall $a_\mu$ calculation. The uncertainty arises in part because the
hadronic cross section lies in the non-perturbative/semi-perturbative regime
where a variety of hadronic resonances lurk. The E989 result makes it more 
implausible that the experimental value is the result of systematic errors.
By combining Fermilab and BNL results, then the quoted discrepancy
is given by\cite{Abi:2021gix}
\be 
a_\mu(Exp)-a_\mu(SM)=(25.1\pm 5.9)\times 10^{-10}
\ee 
corresponding to a $4.2\sigma$ effect. While this latter result was obtained
by comparing experiment to the Theory Initiative value for $a_\mu$, 
we note here that recent lattice evaluations of the HVP contribution to 
$a_\mu$ find a theoretical value which is in accord with the Fermilab/BNL
measured values\cite{Borsanyi:2020mff}.\footnote{In Ref.~\cite{Crivellin:2020zul}, it is
emphasized that a shift in the SM value of HVP to gain accord with 
the $(g-2)_\mu$ measurement would {\it worsen} global fits to EW data.}

An exciting possibility is to account for the muon $g-2$ anomaly by positing 
the existence of weak scale supersymmetry\cite{WSS}, wherein each 
Standard Model (SM) field is elevated to a superfield containing both
fermionic and bosonic components. The additional Higgs fermion (higgsinos)
would destroy the heralded triangle anomaly cancellation within the SM, so
an additional Higgs doublet is needed as well. 
Then, the SM $\mu -\mu-\gamma$ triangle diagrams which contribute to
$a_\mu$ are augmented by  sparticle loops containing $\tchi_{1,2}^--\tnu_\mu$ 
and $\tchi_{1,2,3,4}^0-\tmu_{L,R}$ pairs\cite{Moroi:1995yh}.\footnote{For
some older related references, see \cite{Ellis:1982by,Grifols:1982vx,Barbieri:1982aj,Kosower:1983yw,Yuan:1984ww,Lopez:1993vi,Chattopadhyay:1995ae,Carena:1996qa,Baltz:2001ts,Everett:2001tq,Martin:2001st,Baer:2001kn,Baer:2002gm,Barger:2004mr}.}
The SUSY contribution is given roughly by\cite{Czarnecki:2001pv}
\be
\Delta a_\mu^{\SUSY}\sim \frac{m_\mu^2 \mu M_i\tan\beta}{m_{\SUSY}^4}
\ee
where $\mu$ is the superpotential $\mu$ parameter, $M_i$ is the 
gaugino mass for gauge group $i$, $\tan\beta$ is the ratio of Higgs field
vevs and $m_{\SUSY}$ is a loop average of scalar muons, mu-sneutrinos
and electroweakinos.
By performing a rather general scan over weak scale Minimal 
Supersymmetric Standard Model (MSSM) parameters, then it is found that
in order to explain the muon $g-2$ anomaly, 
the mass of the lightest observable SUSY particles $m_{LOSP}$ must 
be $m_{LOSP}\alt 900$ GeV\cite{Feng:2001tr,Baer:2001kn}. 
This result brings in considerable 
tension with recent LHC search results which so far find no direct evidence for
SUSY particle production at ATLAS or CMS\cite{Canepa:2019hph}.
At present, LHC searches require slepton masses $m_{\tmu_{L,R},\tnu_\mu}\agt 700$ GeV
from 139 fb$^{-1}$ of data at $\sqrt{s}=13$ TeV\cite{Aad:2019vnb,Sirunyan:2020eab}. 
These mass limits are considerably and even entirely relaxed in the case
of degeneracy where $m_{\tmu_{L,R}}\simeq m_{\tchi_1^0}$ (dark matter
coannihilation region).

While it is possible to interpret the presence of the Fermilab/BNL 
$(g-2)_\mu$ anomaly in terms of light smuons and mu-sneutrinos 
along with light electroweakinos, such an explanation seems increasingly 
implausible in light of the global picture of SUSY theory as compared 
with a large array of data from many experiments along with 
theoretical considerations.
\begin{itemize}
\item {\it LHC sparticle search limits:} We have already remarked that
searches for slepton pair production require $m_{\tmu_{L,R}}\agt 700$ GeV
(unless one lives in the degeneracy region where $m_{\tmu_{L,R}}\sim m_{\tz_1}$). 
But also with 139 fb$^{-1}$ 
at $\sqrt{s}=13$ TeV, ATLAS and CMS now require gluinos $m_{\tg}\agt 2.25$
TeV and top squarks $m_{\tst_1}\agt 1.1$ TeV\cite{Canepa:2019hph}. 
While it is certainly possible
that $m_{\tell_{L,R}}\ll m_{\tg},m_{\tst_1}$, such a scenario would imply 
large splitting in the slepton/squark sector. Such splitting seems
especially unlikely when sfermions occupy the 16-dimensional
spinor of $SO(10)$ which provides a beautiful unification of all 
members of each generation.
\item {\it Higgs boson mass:} The LHC measured value of the light
Higgs scalar is $m_h=125.10\pm 0.14$ GeV. 
Such a large mass famously requires TeV-scale top-squarks and 
large mixing\cite{Baer:2011ab} (large trilinear soft terms $A_t$) or
else small mixing but top-squarks in the tens-to-hundreds of TeV, far
beyond any naturalness limits\cite{Arbey:2011ab}. 
Again, a large squark-slepton mass 
splitting would be required to explain both the $(g-2)_\mu$ anomaly and
the Higgs mass.
\item {\it Higgs couplings:} LHC measurements of Higgs boson couplings 
find them to be increasingly SM-like\cite{Aad:2015pla,Sirunyan:2018koj}. In contrast, if SUSY particles
are light, then one expects deviations to occur, especially for
light scalars $H,A$ which would lead to large mixing in the Higgs 
sector\cite{Bae:2015nva}. While SM-like
Higgs couplings can also be explained via alignment\cite{Haber:2018ltt}, 
the decoupling solution seems more plausible.
\item {\it Flavor-changing $B$ decays:} The decay $b\to s\gamma$ is 
particularly interesting in that SM and SUSY loops contribute at 
the same level, so it is a natural place to look for BSM deviations. 
The decay $b\to s\gamma $ takes place via $tW^-$ loops in the SM and via
$\tst_i\tchi_j^-$ and $tH^-$ loops in SUSY \cite{Bertolini:1990if} 
(other SUSY loops also contribute but typically with much smaller 
amplitudes). The SM
value for this decay is found to be \cite{Misiak:2015xwa} ${\rm BF}(b\to
s\gamma )=(3.36\pm 0.23)\times 10^{-4}$ which is to be compared to 
the recent Belle measurement\cite{Belle:2016ufb} that 
${\rm BF}(b\to s\gamma )=(3.01\pm 0.22)\times 10^{-4}$. 
The overlap between theory and experiment are in accord with top-squarks and
charged Higgs bosons in the multi-TeV regime\cite{Baer:2016bwh}.
\item {\it FCNCs from SUSY:} Flavor changing neutral current interactions
(FCNC) are notably absent in the SM due to the GIM 
(Glashow-Iliopoulos-Maiani) mechanism, but generically
ought to be present in SUSY extensions\cite{Gabbiani:1996hi} unless there is 1. flavor universality in the scalar sector or 2. alignment between quark and squark mass 
matrices\cite{Nir:1993mx} or 
3. decoupling due to sfermions in the 10-100 TeV 
regime\cite{Dine:1993np,Cohen:1996vb}.
In generic gravity-mediated SUSY breaking models, non-universality is the rule
while very special circumstances must be realized for alignment. Meanwhile,
the decoupling solution is found to be consistent with the 
{\it electroweak naturalness measure} $\Delta_{\EW}$ since first/second
generation contributions to the weak scale are all suppressed due to the 
small Yukawa couplings. Models such as gauge-mediated SUSY breaking 
(GMSB) and minimal anomaly-mediation (mAMSB) that do conserve flavor 
seem excluded in the former case by the Higgs mass constraint and 
naturalness\cite{Baer:2014ica}
and in the latter case by the Higgs mass constraint, naturalness and 
the presence of a wino-like lightest SUSY particle (LSP)\cite{Baer:2018hwa}.\footnote{
Note that {\it natural} AMSB models which include a bulk trilinear
term and non-universal bulk Higgs masses are consistent with naturalness, 
Higgs mass and dark matter constraints\cite{Baer:2014ica}. 
They have a higgsino rather than a wino as LSP.}  
\item {\it SUSY $CP$ violation:} Likewise, SUSY soft terms are expected to
be complex in general leading to contributions to low energy CP violating 
processes\cite{Gabbiani:1996hi}. 
The offending CP violating processes are all suppressed by 
a scalar mass decoupling solution with scalar masses in the several-TeV regime.
\item {\it Cosmological gravitino problem:} In gravity-mediation, the 
gravitino mass $m_{3/2}$ sets the mass scale for the soft breaking terms,
and hence the sparticle masses. But if the gravitino is too light-- 
below the TeV scale-- then it can be long-lived.
Then gravitinos produced thermally in the early universe will decay after
the culmination of Big Bang nucleosynthesis (BBN) and disrupt the successful
BBN predictions of the hot Big Bang model. 
Increasing $m_{3/2}$ above about $m_{3/2}\agt 5$ TeV 
is usually sufficient 
in order to shorten the gravitino lifetime so 
that it decays before BBN\cite{Khlopov:1984pf,Kawasaki:2008qe}.
\item {\it Cosmological moduli problem:} Light moduli fields-- 
gravitationally coupled string remnant scalar fields from the compactification
of extra dimensions-- may also exist (where by light we mean of order the 
soft SUSY breaking terms). 
Any light modulus could be produced in the early universe due to coherent
scalar field oscillations, and if they are too light-- typically below about
30 TeV-- then like gravitinos, they would disrupt the successful light element
production in the early universe\cite{Linde:1996cx}.
\item {\it Naturalness allows heavy first/second generation matter scalars:} 
Heavy scalar fields in SUSY models seemingly violate older 
notions of naturalness which was the motivation for sparticles 
at or around the weak scale 
(which would provide large contributions to $(g-2)_\mu$).
However, naturalness based on tuning of parameters $p_i$,
$\Delta =max_i|\partial\log m_Z^2/\partial\log p_i|$, was typically based on
numerous free soft term parameters in the low energy SUSY effective field theory
(EFT). In string theory, in our universe, the soft terms are expected to be 
all correlated and calculable in the case of stabilized moduli.
Alternatively, naturalness evaluations requiring logarithmic radiative 
corrections to the light Higgs mass to be small would oversimplify and neglect
that the SUSY radiative corrections depend on the Higgs soft mass itself
(radiative corrections not independent of tree-level masses) and further 
would have to extrapolate radiative corrections across the electroweak 
phase transition (which is not the case in the SM). 
By using the more conservative $\Delta_{\EW}$ measure\cite{Baer:2012up,Baer:2012cf}, then 
few-TeV scale top-squarks and $10-50$ TeV first/second generation matter 
scalars are allowed.
\item {\it Weak scale SUSY from the string landscape:} The presence of a vast 
landscape of metastable string vacua now seems inescapable\cite{Douglas:2006es}. The idea
has some experimental support in that it allows for a solution to the
cosmological constant problem and indeed even predicts its value to within 
a factor of a few which is confirmed by dark energy measurements. 
The landscape is also expected to favor a power-law draw to large soft terms
which are limited by the ABDS\cite{Agrawal:1998xa} anthropic window of not-to-large a value of
the weak scale in pocket universes which are compatible with atoms as we 
know them (atomic principle). In this case, statistical predictions can be
made for Higgs and superparticle masses. Indeed, it is found that the landscape
statistically favors a light Higgs scalar with mass $m_h\sim 125$ GeV with
sparticles (save light higgsinos) generally well beyond LHC search 
limits\cite{Baer:2020kwz}. 
The landscape pulls first/second generation matter scalar masses
into the $10-50$ TeV regime. This would lead to only tiny SUSY contributions 
to $(g-2)_\mu$.
\end{itemize}

Thus, to summarize the above bullet points, there are a variety of 
reasons from both theory and experiment to expect that nature is 
supersymmetric, but with SUSY particles, save light higgsinos, 
in the $\sim(1-50)$ TeV range.
 
In this paper, we examine whether it is possible to reconcile the 
latest measurement of the muon $(g-2)_\mu$ anomaly with weak scale 
supersymmetry in light of recent LHC search limits and while requiring
weak scale naturalness (in that unnatural models, while logically possible, 
are highly implausible). 
We also examine $(g-2)_\mu$ in the recent context of 
the string theory landscape. While the above bullet points indicate
TeV-scale soft terms, the muon $(g-2)_\mu$ anomaly requires sub-TeV
muonic scalars\cite{Feng:2001tr}. These requirements were reconciled
long ago within the context of normal scalar mass hierarchy models wherein
the first two generations of scalars were much lighter than third 
generation scalars\cite{Baer:2004xx}.\footnote{see also Ref. \cite{Yamaguchi:2016oqz}.} 
Such a situation is possible within the three-extra-parameter 
non-universal Higgs model (NUHM3)\cite{nuhm2,nuhm22,nuhm23,nuhm24,nuhm25,nuhm26}, which also allows for 
EW naturalness and which we adopt for much of our analysis.
The NUHM3 model is embedded in the computer code Isajet 7.88\cite{isajet} 
which we use for sparticle spectra generation and for evaluation
of $\Delta a_\mu^{\SUSY}$.
%
%

A number of other papers have recently appeared regarding a SUSY
interpretation of the Fermilab/BNL $(g-2)_\mu$ anomaly\cite{Endo:2021zal,Iwamoto:2021aaf,Gu:2021mjd,VanBeekveld:2021tgn,Yin:2021mls,Wang:2021bcx,Abdughani:2021pdc,Cao:2021tuh,Chakraborti:2021dli,Ibe:2021cvf,Cox:2021gqq,Han:2021ify,Heinemeyer:2021zpc,Baum:2021qzx,Zhang:2021gun,Ahmed:2021htr,Athron:2021iuf,Aboubrahim:2021rwz,Chakraborti:2021bmv}.

\section{A benchmark SUSY point which can explain $(g-2)_\mu$ anomaly}
\label{sec:bm}

In Table \ref{tab:bm}, we show the spectra from a NUHM3 benchmark 
point which fulfills 1. LHC Higgs mass and sparticle limits, 
2. is natural ($\Delta_{\EW}=29.9$) 3. provides sufficient
contribution to explain the Fermilab/BNL $(g-2)_\mu$ anomaly
and 4. is dark matter allowed.\footnote{This point is generated using
a modifed version of Isajet 7.78\cite{isajet} where a common scale choice
$Q^2=m_{\tst_1}m_{\tst_2}$ is used to evaluate the Higgs mass and
sparticle self-energies\cite{pbmz}. In an earlier version, the highly split
spectrum lead to unstable radiative corrections due to very large
logs in the self-energy corrections that arose due to the strong disparity
in mass scales. We thank C. Wagner for his sharp eyes.} 
It has a higgsino-like LSP so that
there is a deficit of neutralino dark matter. This leaves room for
axions (thus two dark matter particles, axions and WIMPs) which result 
from an assumed solution to the strong CP problem\cite{Baer:2011hx}.
This benchmark point has $m_{\tg}\sim 2340$ GeV and $m_{\tst_1}\sim 1348$ GeV
but has a spectrum of light sleptons. The unusual pattern of slepton masses
occurs due to a strong influence of the SUSY renormalization group
$S$-term (see p. 206 of Ref. \cite{WSS}) which is zero in models 
with scalar mass universality but is
large in this case due to the large splitting between $m_{H_u}^2$
and $m_{H_d}^2$. This RGE term boosts up the right-slepton mass to 
over a TeV while the left-sleptons have masses below 300 GeV. 
In this case, the $\mu$ parameter is large enough that the mu-sneutrino 
lies in the quasi-degeneracy region of the LHC slepton search exclusion plot
where visible slepton decay products become soft enough so as to evade the 
bounds. Furthermore, the dominant decay mode is actually to invisibles: 
$\tnu_{\mu L}\to\nu_\mu\tz_1$.
The left smuon lies precisely on the edge of the LHC slepton 
excluded region. However, its dominant decay mode is\cite{isajet}
$\tmu_L\to \tchi_1^-\nu_\mu$ (where the $\tchi_1^-$ visible decay products are
extremely soft due to the small mass gap $m_{\tw_1}-m_{\tz_1}\sim 13$ GeV)
which is radically different from the assumed simplified models used in the
ATLAS/CMS exclusion plots.
Thus, the point should at present be allowed by LHC slepton searches. 

The point is perhaps artificial in that the first two generations are taken 
as light and degenerate whilst the third generation is far heavier.
This point is an example of a normal scalar mass 
hierarchy model\cite{Baer:2004xx}
which was introduced to reconcile $\Delta a_\mu^{\SUSY}$ 
(which requires light smuons) with the near match in theory-experiment for
$b\to s\gamma$ decays (which requires heavier top-squarks to suppress 
non-standard contributions). Nowadays, it also helps to avoid 
LHC top-squark search limits and to explain the large 
light Higgs mass $m_h$ which in SUSY requires TeV-scale highly mixed
stops to lift its mass to $\sim 125$ GeV. Naively, one might expect
such a point to lead to detectable amounts of 
lepton-flavor-violation (LFV)\cite{Lindner:2016bgg,Baer:2019xug}.

\begin{table}\centering
\begin{tabular}{lc}
\hline
parameter & $NUHM3$ \\
\hline
$m_0(1,2)$      & 368 \\
$m_0(3)$      & 2955 \\
$m_{1/2}$   & 1055  \\
$A_0$      & -4370 \\
$\tan\beta$    & 26  \\
$\mu$          & 230 \\
$m_A$          & 5365 \\
\hline
$m_{\tg}$   & 2341.6 \\
$m_{\tu_L}$ & 2148.8\\
$m_{\tu_R}$ & 1848.6 \\
$m_{\te_R},\ m_{\tmu_R}$ & 1124.9 \\
$m_{\te_L},\ m_{\tmu_L}$ & 297.6 \\
$m_{\tnu_{eL}},\ m_{\tnu_{\mu L}}$ & 284.9 \\
$m_{\tst_1}$ & 1347.5 \\
$m_{\tst_2}$ & 2613.0 \\
$m_{\tb_1}$ & 2647.6 \\
$m_{\tb_2}$ & 3192.0 \\
$m_{\ttau_1}$ & 2648.3 \\
$m_{\ttau_2}$ & 2704.0 \\
$m_{\tnu_{\tau}}$ & 2695.6 \\
$m_{\tw_2}$ & -864.7 \\
$m_{\tw_1}$ & -240.0 \\
$m_{\tz_4}$ & -875.4 \\ 
$m_{\tz_3}$ & -460.7 \\ 
$m_{\tz_2}$ & 239.7 \\ 
$m_{\tz_1}$ & -227.1 \\ 
$m_h$       & 124.1 \\ 
\hline
$\Delta a_\mu^{\SUSY}$ & $18.6\times 10^{-10}$ \\
$BF(b\to s\gamma)\times 10^4$ & 2.8 \\
$BF(B_s\to \mu^+\mu^-)\times 10^9$ & 3.8 \\
$\Omega_{\tz_1}^{std}h^2$ & 0.015  \\
$\sigma^{SI}(\tz_1, p)$ (pb) & $2.7\times10^{-9}$ \\
$\sigma^{SD}(\tz_1 p)$ (pb) & $5.2\times10^{-5}$ \\
$\langle\sigma v\rangle |_{v\to 0}$  (cm$^3$/sec)  & $1.6\times10^{-25}$ \\
$\Delta_{\rm EW}$ & 29.9 \\
\hline
\end{tabular}
\caption{Input parameters and masses in~GeV units
for a natural NUHM3 SUSY benchmark point
with a normal scalar mass hierarchy and with $m_t=173.2$ GeV. 
}
\label{tab:bm}
\end{table}

\section{$(g-2)_\mu$ from natural SUSY}
\label{sec:nat}

In this Section, we show the expected value of $\Delta a_\mu^{\SUSY}$ from 
SUSY with electroweak naturalness $\Delta_{\EW}\alt 30$. 
This measure requires the various independent SUSY contributions to the 
weak scale to be comparable to or less than the weak scale: 
\be
m_Z^2/2=\frac{m_{H_d}^2+\Sigma_d^d-(m_{H_u}^2+\Sigma_u^u)\tan^2\beta}{\tan^2\beta -1}-\mu^2\simeq -m_{H_u}^2-\Sigma_u^u(\tst_{1,2})-\mu^2
\label{eq:mzs}
\ee
where
\be
\Delta_{EW}=max|{\rm terms\ on\ right\ hand\ side\ of\ Eq.~\ref{eq:mzs}}|/(m_Z^2/2) .
\ee
The $\Sigma_d^d$ and $\Sigma_u^u$ contain over 40 1-loop corrections which are 
listed in Ref. \cite{Baer:2012cf}.

To proceed, we adopt the two- and three-extra parameter
non-universal Higgs mass models which allow the presence of small 
higgsino masses $\mu$ (typical SUSY spectrum generators will tune the value of
$\mu$ to exactly the right large unnatural value so as to ensure a 
$Z$-boson mass of $m_Z=91.2$ GeV).
The NUHM2 parameter space is given by
\be
m_0,\ m_{1/2},\ A_0,\ \tan\beta,\ m_{H_u},\ m_{H_d}
\ee
where the scalar potential minimization conditions allow one to 
trade $m_{H_u}$ and $m_{H_d}$ for the more convenient weak scale
variables $\mu$ and $m_A$. The NUHM3 model goes a step further
and allows a common mass $m_0(1,2)$ for first/second generation
matter scalars and an independent mass $m_0(3)$ for third generation 
scalars.

We next scan over the following NUHM2,3 parameter space.
\bea
m_0&:&\ 0-13\ {\rm TeV}\nonumber \\
m_{1/2}&:&\ 0-3\ {\rm TeV},\nonumber \\ 
-A_0&:&\ 0-20\ {\rm TeV},\nonumber \\ 
m_A&:&\ 0-10\ {\rm TeV},\nonumber \\
\mu &:&\ 100-360\ {\rm GeV}\nonumber\\
\tan\beta &:&\ 3-60. \nonumber
\eea
For NUHM3, we additionally scan on $m_0(1,2):\ 0-55$ TeV.
The scan upper limits are taken beyond the upper limits imposed by
naturalness ($\Delta_{\EW}<30$) so are not artificial.
For surviving scan points, we only require an appropriate breakdown
of electroweak symmetry and the Higgs mass to lie within its
measured range $123\ {\rm GeV}<m_h<127$ GeV (allowing for $\pm2$ GeV
theory uncertainty in the Higgs mass calculation).

\begin{figure}[tbh]
\begin{center}
\includegraphics[height=0.4\textheight]{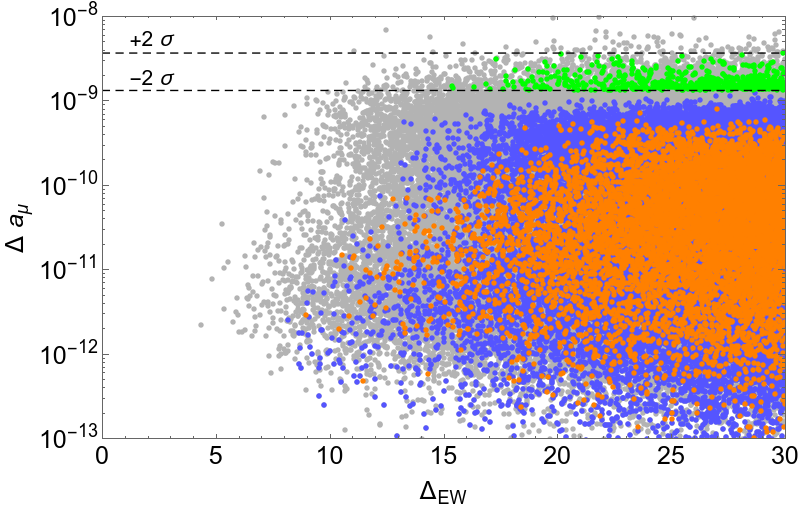}
\caption{Value of $\Delta a_\mu^{\SUSY}$ from natural SUSY vs. 
naturalness measure $\Delta_{\EW}$. 
Gray dots are from NUHM2,3 models with $123\ {\rm GeV}<m_h<127$ GeV but which violate
LHC search limits. 
Orange (NUHM2) and blue (NUHM3) points are LHC-allowed.
Green (NUHM3) points are LHC-allowed, natural, have the right Higgs mass 
and occupy the $(g-2)_\mu$ $2\sigma$ band.
\label{fig:dew}}
\end{center}
\end{figure}

Our first results are shown in Fig. \ref{fig:dew}, where we show our NUHM2,3 
scan points in the $\Delta_{\EW}$ vs. $\Delta a_\mu^{\SUSY}$ plane. 
The region between the dashed lines denotes the combined
BNL/Fermilab measurement (at $2\sigma$ significance). 
The gray points denote NUHM2,3 points with naturalness $\Delta_{\EW}<30$  
and light Higgs boson mass within the range $m_h:\ 123-127$ GeV 
but which otherwise are excluded by LHC search limits. 
The orange points are from the NUHM2 model which forces degeneracy
between the first/second generation matter scalars (second generation 
scalars mainly contribute to 
the $(g-2)_\mu$ anomaly) and third generation scalars (whose
contributions $\Sigma_u^u(\tst_{1,2})$ impact upon naturalness). 
The NUHM3 points, which relax this degeneracy, are denoted as blue. 
Both the orange NUHM2 points and the blue NUHM3 points satisfy in addition 
LHC search constraints that
\bi
\item $m_{\tg}>2.25$\ {\rm TeV},
\item $m_{\tst_1}>1.1$\ {\rm TeV}\ {\rm and}
\item $m_{\tell}>700$\ {\rm GeV}.
\ei
Finally, the green points denote NUHM3 points which obey the above 
LHC constraints {\it except the last one}, where $m_{\tell}<700$ GeV is 
allowed if one lives in the
quasi-degeneracy region where $m_{\tell}\sim m_{\tz_1}$ and/or if the 
slepton branching fractions deviate significantly from the ATLAS/CMS simplified model assumptions. 
These points, like our benchmark point, can allow the anomalous muon 
magnetic moment to occupy the $2\sigma$ band while still respecting 
electroweak naturalness
and LHC Higgs and sparticle mass limits. 

While many of the gray points populate the BNL/Fermilab band, they are all
excluded by LHC search limits.
Also, the bulk of blue and orange points have $\Delta a_\mu^{\SUSY}$ values well below the 
anomalous region.
When the Higgs mass constraint is applied to NUHM2 points, then sleptons are too heavy 
to generate the preferred anomaly band, so the orange points all lie well below 
the $(g-2)_\mu$ $2\sigma$ band.
However, if we relax generational degeneracy (blue and green points), 
then we can allow for the normal scalar mass hierarchy model.
In this case, the blue points almost reach the BNL/Fermilab $2\sigma$ band; 
the green points are then the subset of NUHM3 points which fulfill all LHC bounds, 
the naturalness condition and explain the BNL/Fermilab $(g-2)_\mu$ anomaly.

\begin{figure}[tbh]
\begin{center}
\includegraphics[height=0.4\textheight]{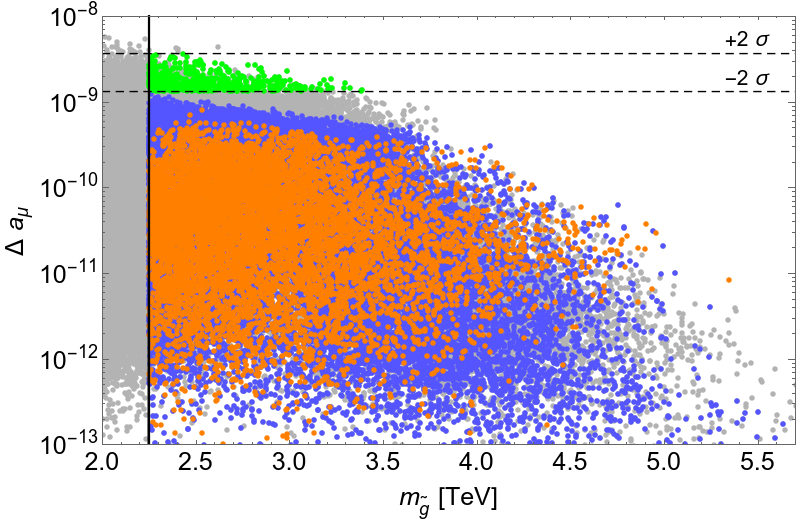}
\caption{Value of $\Delta a_\mu^{\SUSY}$ from natural SUSY vs. 
gluino mass $m_{\tg}$. 
Gray dots are from NUHM2,3 models with $123\ {\rm GeV}<m_h<127$ GeV 
but which violate LHC search limits. 
Orange (NUHM2) and blue (NUHM3) points are LHC-allowed.
Green (NUHM3) points are LHC-allowed, natural, have $m_h:\ 123-127$ GeV
and occupy the $(g-2)_\mu$ $2\sigma$ band.
\label{fig:mgl}}
\end{center}
\end{figure}

In Fig. \ref{fig:mgl}, we show the distribution of scan points 
in the $m_{\tg}$ vs. $\Delta a_\mu^{\SUSY}$ plane. 
The color coding is as in Fig. \ref{fig:dew} 
The ATLAS/CMS limit that $m_{\tg}\agt 2.25$ TeV is denoted by the vertical line. 
In order to obtain $m_h\sim 125$ GeV within the orange NUHM2 points,  
the soft breaking stop mass terms must be in the several TeV regime and 
also have large mixing from a large $A_0$ parameter. 
The first of these tends to pull the sleptons up to high masses which suppress
$\Delta a_\mu^{\SUSY}$. 
This effect can be avoided by allowing non-universal generations $m_0(1,2)\ne m_0(3)$
whereby heavier stops and $m_h\sim 125$ GeV can co-exist with lighter sleptons,
as in the LHC-allowed blue and green points.
Once the LHC sparticle and Higgs mass constraints are imposed, along with naturalness, 
then we see that the green points which lie within the
BNL/Fermilab anomalous region allow for gluino masses up to 3.5 TeV.
This is in contrast to naturalness-only upper bounds on $m_{\tg}$
which allow for $m_{\tg}$ to range up to $\sim 6$ TeV.

\begin{figure}[tbh]
\begin{center}
\includegraphics[height=0.4\textheight]{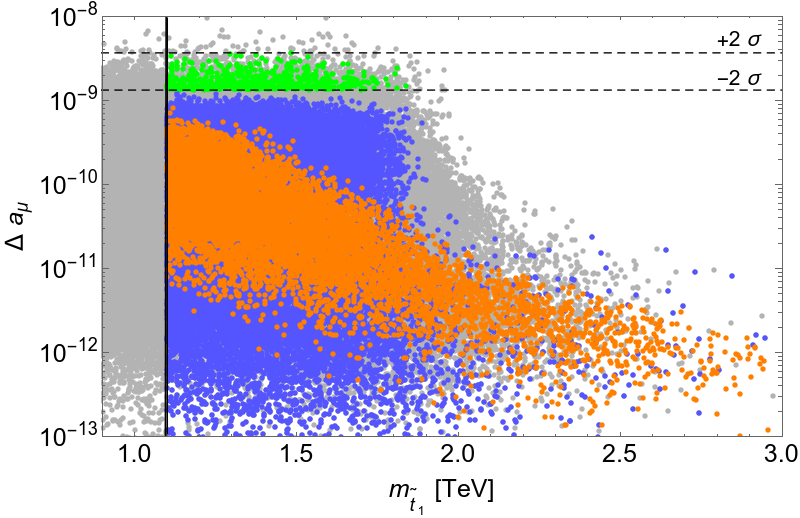}
\caption{Value of $\Delta a_\mu^{\SUSY}$ from natural SUSY vs. 
light top-squark mass $m_{\tst_1}$. 
Gray dots are from NUHM2,3 models with $123\ {\rm GeV}<m_h<127$ GeV 
but which violate LHC search limits. 
Orange (NUHM2) and blue (NUHM3) points are LHC-allowed.
Green (NUHM3) points are LHC-allowed, natural, have the right Higgs mass 
and occupy the $(g-2)_\mu$ $2\sigma$ band.
\label{fig:mt1}}
\end{center}
\end{figure}

In Fig. \ref{fig:mt1}, we show $\Delta a_\mu^{\SUSY}$ vs. light stop mass $m_{\tst_1}$.
The color-coding is again as in Fig. \ref{fig:dew}. 
The EW natural points with the allowed $m_h$ value occur for $m_{\tst_1}\alt 3$ TeV--
well beyond HL-LHC projected reach values.
For the orange NUHM2 points with universality, then the larger $m_{\tst_1}$ becomes, the lower
is the allowed $\Delta a_\mu^{SUSY}$.
But for the NUHM3 points which allow for $m_0(1,2)<m_0(3)$, then green points which fulfill
naturalness, LHC Higgs and sparticle search bound are found which can explain
the $(g-2)_\mu$ anomaly.
These points actually have a much reduced upper bound on $m_{\tst_1}$ where
$m_{\tst_1}\alt 1.7$ TeV. This is good news for HL-LHC top-squark 
searches since their projected reach with 3000 fb$^{-1}$ is to 
$m_{\tst_1}\sim 1.7$ TeV\cite{CidVidal:2018eel}.

\begin{figure}[tbh]
\begin{center}
\includegraphics[height=0.4\textheight]{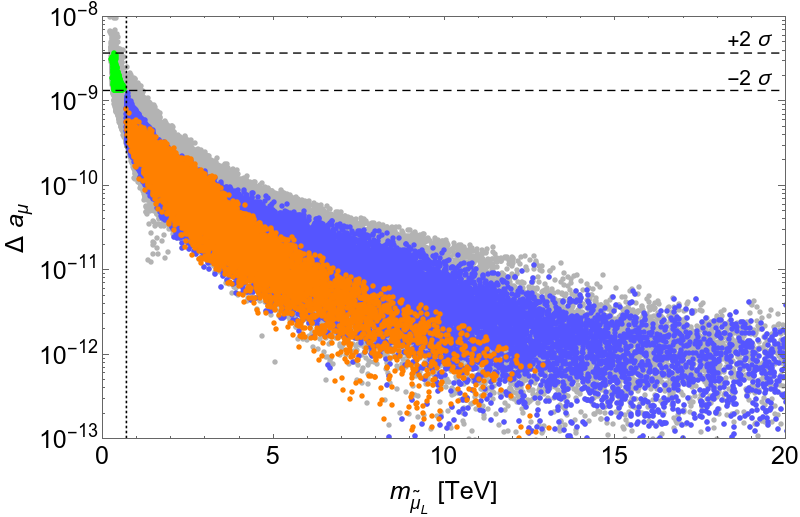}
\caption{Value of $\Delta a_\mu^{\SUSY}$ from natural SUSY vs. 
left-smuon mass $m_{{\tmu_ L}}$. 
Gray dots are from NUHM2,3 models with $123\ {\rm GeV}<m_h<127$ GeV 
but which violate LHC search limits. 
Orange (NUHM2) and blue (NUHM3) points are LHC-allowed.
Green (NUHM3) points are LHC-allowed, natural, have the right Higgs mass 
and occupy the $(g-2)_\mu$ $2\sigma$ band.
\label{fig:msnu}}
\end{center}
\end{figure}

In Fig. \ref{fig:msnu}, we show the distribution of $\Delta a_\mu^{\SUSY}$
versus the left smuon mass $m_{{\tmu_ L}}$. 
The vertical line denotes the LHC limit that $m_{\tell}>700$ for the
non-degeneracy region and assuming the simplified model decay $\tell\to\ell\tz_1$.
The orange NUHM2 points almost always have far heavier slepton masses $m_{\tell}\agt 700$ GeV.
Meanwhile, the blue and green NUHM3 points allow for much lighter sleptons while
fulfilling LHC search constraints. The green points are still LHC-allowed 
because, even with $m_{\tell}<700$ GeV, they can lie in the quasi-degeneracy 
region where $m_{\tell}\sim m_{\tz_1}$ or they have decay branching fractions
which deviate significantly from LHC simplified model assumptions. 
Overall, we see the well-known result
that the $(g-2)_\mu$ anomaly can only be explained in models with 
$m_{{\tmu_L}}\alt 900$ GeV 
(as shown long ago by Feng and Matchev\cite{Feng:2001tr}).
The bulk of models have far higher smuon and mu-sneutrino masses and 
consequently much tinier contributions to $\Delta a_\mu^{\SUSY}$.

\section{Expected value of $\Delta a_\mu^{\SUSY}$ from the landscape}
\label{sec:results}

Next we move to the predictions of string theory for the muon 
anomalous magnetic moment. We will appeal to the string theory
landscape\cite{Susskind:2003kw}, wherein string flux compactifications\cite{Douglas:2006es} lead to an 
enormous number of metastable string vacua, say $10^{500}-10^{272,000}$,
each with different moduli vevs and hence each with varying $4-d$ 
laws of physics. 
The existence of the landscape is by now widely accepted in the
string community and indeed it provides the only plausible
solution to the cosmological constant problem via Weinberg's 
anthropic reasoning\cite{Weinberg:1987dv}.

Next, we will assume a fertile patch of $4-d$ landscape of vacua 
including the MSSM as the low energy EFT. 
For spontaneous SUSY breaking in such a patch via 
complex-valued $F$-terms $F_i$ and real values $D$-terms $D_{\alpha}$, 
then the overall SUSY breaking scale $m_{soft}$ is 
statistically favored for large values as a power-law\cite{Douglas:2004qg}
$f_{\SUSY}\sim m_{soft}^n$ where the exponent $n=2n_F+n_D-1$. 
Then, even for the textbook case of SUSY breaking via a single $F$-term,
one expects an $n=1$ draw to large soft breaking. However, the draw to 
large soft breaking must be tempered by the anthropic requirement
of not-to-large a value of the weak scale in each pocket universe of
our fertile patch. The Agrawal {\it et al.}\cite{Agrawal:1998xa} (ABDS) anthropic window 
requires that the pocket universe weak scale be not-too-far
from the measured weak scale value in our universe: 
$m_{weak}^{\PU}<(2-5)m_{weak}^{\OU}$. 
For too large a value of the weak scale, then complex nuclei
are no longer allowed and atoms as we know them would not arise 
(atomic principle).
To be specific, we take
$m_{weak}^{\PU}< 4 m_{weak}^{\OU}$ which corresponds to EW finetuning 
measure $\Delta_{\EW}<30$. 
Under such a scenario, one can make statistical predictions for Higgs 
and sparticle masses. 
For $n\sim 1-4$\cite{Baer:2017uvn} or even a log distribution\cite{Baer:2020dri} of soft parameters, one gains
a peak for $m_h\sim 125$ GeV with sparticle masses lifted beyond present 
LHC search limits. 

\begin{figure}[tbh]
\begin{center}
\includegraphics[height=0.4\textheight]{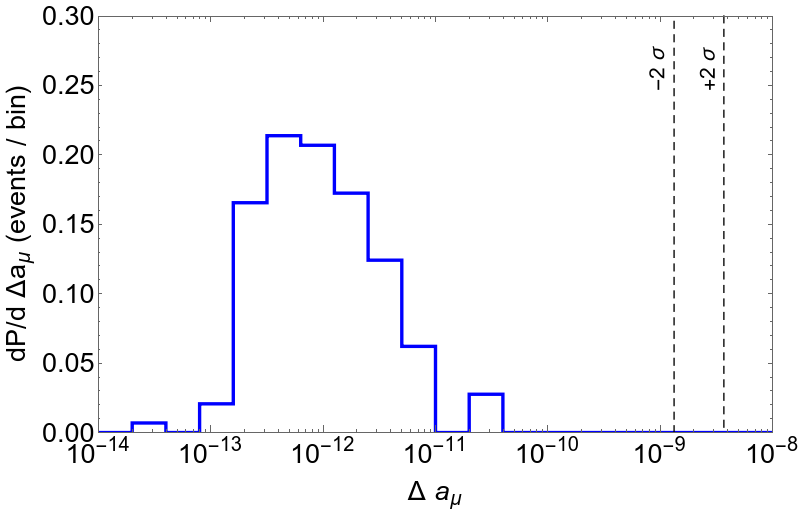}
\caption{Distribution of $\Delta a_\mu^{\SUSY}$ values from an $n=1$ 
power-law draw for soft terms in the string theory 
landscape where NUHM3 is the low energy EFT.
We require the pocket universe weak scale to lie within the ABDS window
with $m_{weak}^{\PU}<4m_{weak}^{\OU}$.
\label{fig:damu}}
\end{center}
\end{figure}

In Fig. \ref{fig:damu}, we show the string landscape prediction for 
$\Delta a_\mu^{\SUSY}$ assuming that the low energy EFT is the NUHM3 model.
The different soft terms should scan independently of each other\cite{Baer:2020vad}.
The first/second generation soft terms $m_0(1,2)$ are pulled to
very large values $\sim 10-50$ TeV since they contribute to 
the weak scale via tiny Yukawa-suppressed terms or via two-loop RGEs.
Then the smuons become correspondingly very heavy and the prediction 
is for a tiny SUSY contribution to the $(g-2)_\mu$ anomaly.
From the Figure, we see a peak probability distribution
around $\Delta a_\mu\sim 5\times 10^{-13}$ with almost no probability
extending to the anomaly region. While this is bad for explaining the
$(g-2)_\mu$ anomaly, it does provide a mixed decoupling/quasi-degeneracy
solution to the SUSY flavor and $CP$ problems\cite{Baer:2019zfl}.

\section{Conclusions}
\label{sec:conclude}

In this paper, we have examined the possibility of natural SUSY and 
also landscape SUSY to explain the BNL/Fermilab $(g-2)_\mu$ anomaly.
We focus on natural SUSY models since they fufill the plausibility
requirement of also providing a natural explanation for why the weak 
scale $m_{W,Z,h}\sim 100$ GeV and not in the TeV regime 
(thus solving the Little Hierarchy Problem).
We presented an example in Table \ref{tab:bm} of a natural SUSY point 
which explains the $(g-2)_\mu$ anomaly while respecting LHC Higgs mass
and sparticle search limits, while remaining natural.
It is characterized by a normal scalar mass hierarchy where
$m_0(1,2)\ll m_0(3)$ leads to light sleptons. 
The light sleptons are still LHC-allowed because either they
lie in the quasi-degeneracy region
where $m(slepton)\sim m(LSP)$ and/or their branching fractions 
deviate significantly from the 
ATLAS/CMS simplified model assumption.

Otherwise, for the natural models, it is very hard to explain the anomaly while
simultaneously requiring the calculated value of $m_h\sim 123-127$ GeV
and also obeying LHC constraints that $m_{\tg}>2.25$ TeV, 
$m_{\tst_1}>1.1$ TeV and $m_{\tell}>700$ GeV.\footnote{For a related analysis
of LHC constraints and $\Delta a_\mu^{\SUSY}$, see Ref. \cite{Hagiwara:2017lse}.}
Basically, the LHC Higgs and sparticle mass constraints favor multi-TeV
soft terms which then also should pull the smuon sector to large 
masses which then suppresses the $\Delta a_\mu^{\SUSY}$ anomaly contribution.
One can in principle explain the anomaly within SUSY by working in the pMSSM
and taking light smuons and light LSP while the remaining sparticles 
are in the TeV-and-beyond region. 
But such models with uncorrelated spectra may be pathological
(in that they disregard the RGEs and hence may miss out on
gauge coupling unification and radiative electroweak symmetry breaking and
a light Higgs mass determined by correlations between parameters) 
so it is unclear if they should be taken seriously. 

By moving to landscape SUSY, matters get even worse. The landscape is 
expected to pull soft terms to large values via a power-law or 
logarithmic draw until they start providing too large of 
contributions to the weak scale. 
This picture predicts a Higgs mass $m_h\sim 125$ GeV and sparticles beyond 
present LHC bounds.
But since first/second generation sfermions
contribute to the weak scale only via Yukawa suppressed terms or
two-loop RGE contributions, their masses get pulled into the $10-50$ TeV regime
which results in only tiny contributions to the $(g-2)_\mu$ anomaly.

By adopting the big picture, LHC Higgs and sparticle mass limits coupled 
with naturalness make an interpretation within weak scale SUSY seem
rather implausible (but not impossible, as shown by our benchmark point
and the green points of Sec. \ref{sec:nat}). 
Another possible resolution of the BNL/Fermilab $(g-2)_\mu$ anomaly 
which allows for decoupled sleptons
is that perhaps the lattice calculations\cite{Borsanyi:2020mff}--
which lead to agreement between SM theory and experiment--
are correct and that perhaps strong interaction effects disrupt the
dispersive techniques used to calculate the HVP 
which require detailed knowledge of
$e^+e^-\to hadrons$ data at or below the QCD confinement scale.

{\it Acknowledgements:} 

We thank X. Tata for valuable discussions.
This material is based upon work supported by the U.S. Department of Energy, 
Office of Science, Office of High Energy Physics under Award Number DE-SC-0009956 and U.S. Department of Energy (DoE) Grant DE-SC-0017647. 

The computing for this project was performed at the OU Supercomputing Center 
for Education \& Research (OSCER) at the University of Oklahoma (OU).


\bibliography{amu}

\begin{thebibliography}{10}
\expandafter\ifx\csname url\endcsname\relax
  \def\url#1{\texttt{#1}}\fi
\expandafter\ifx\csname urlprefix\endcsname\relax\def\urlprefix{URL }\fi
\expandafter\ifx\csname href\endcsname\relax
  \def\href#1#2{#2} \def\path#1{#1}\fi

\bibitem{Abi:2021gix}
B.~Abi, et~al., {Measurement of the Positive Muon Anomalous Magnetic Moment to
  0.46 ppm}, Phys. Rev. Lett. 126 (2021) 141801.
\newblock \href {http://arxiv.org/abs/2104.03281} {\path{arXiv:2104.03281}},
  \href {http://dx.doi.org/10.1103/PhysRevLett.126.141801}
  {\path{doi:10.1103/PhysRevLett.126.141801}}.

\bibitem{Bennett:2004pv}
G.~W. Bennett, et~al., {Measurement of the negative muon anomalous magnetic
  moment to 0.7 ppm}, Phys. Rev. Lett. 92 (2004) 161802.
\newblock \href {http://arxiv.org/abs/hep-ex/0401008}
  {\path{arXiv:hep-ex/0401008}}, \href
  {http://dx.doi.org/10.1103/PhysRevLett.92.161802}
  {\path{doi:10.1103/PhysRevLett.92.161802}}.

\bibitem{Aoyama:2020ynm}
T.~Aoyama, et~al., {The anomalous magnetic moment of the muon in the Standard
  Model}, Phys. Rept. 887 (2020) 1--166.
\newblock \href {http://arxiv.org/abs/2006.04822} {\path{arXiv:2006.04822}},
  \href {http://dx.doi.org/10.1016/j.physrep.2020.07.006}
  {\path{doi:10.1016/j.physrep.2020.07.006}}.

\bibitem{Davier:2019can}
M.~Davier, A.~Hoecker, B.~Malaescu, Z.~Zhang, {A new evaluation of the hadronic
  vacuum polarisation contributions to the muon anomalous magnetic moment and
  to $\mathbf{\boldsymbol\alpha(m_Z^2)}$}, Eur. Phys. J. C 80~(3) (2020) 241,
  [Erratum: Eur.Phys.J.C 80, 410 (2020)].
\newblock \href {http://arxiv.org/abs/1908.00921} {\path{arXiv:1908.00921}},
  \href {http://dx.doi.org/10.1140/epjc/s10052-020-7792-2}
  {\path{doi:10.1140/epjc/s10052-020-7792-2}}.

\bibitem{Borsanyi:2020mff}
S.~Borsanyi, et~al., {Leading hadronic contribution to the muon 2 magnetic
  moment from lattice QCD}\href {http://arxiv.org/abs/2002.12347}
  {\path{arXiv:2002.12347}}, \href
  {http://dx.doi.org/10.1038/s41586-021-03418-1}
  {\path{doi:10.1038/s41586-021-03418-1}}.

\bibitem{Crivellin:2020zul}
A.~Crivellin, M.~Hoferichter, C.~A. Manzari, M.~Montull, {Hadronic Vacuum
  Polarization: $(g-2)_\mu$ versus Global Electroweak Fits}, Phys. Rev. Lett.
  125~(9) (2020) 091801.
\newblock \href {http://arxiv.org/abs/2003.04886} {\path{arXiv:2003.04886}},
  \href {http://dx.doi.org/10.1103/PhysRevLett.125.091801}
  {\path{doi:10.1103/PhysRevLett.125.091801}}.

\bibitem{WSS}
H.~Baer, X.~Tata, {Weak scale supersymmetry: From superfields to scattering
  events}, Cambridge University Press, 2006.

\bibitem{Moroi:1995yh}
T.~Moroi, {The Muon anomalous magnetic dipole moment in the minimal
  supersymmetric standard model}, Phys. Rev. D 53 (1996) 6565--6575, [Erratum:
  Phys.Rev.D 56, 4424 (1997)].
\newblock \href {http://arxiv.org/abs/hep-ph/9512396}
  {\path{arXiv:hep-ph/9512396}}, \href
  {http://dx.doi.org/10.1103/PhysRevD.53.6565}
  {\path{doi:10.1103/PhysRevD.53.6565}}.

\bibitem{Ellis:1982by}
J.~R. Ellis, J.~S. Hagelin, D.~V. Nanopoulos, {Spin 0 Leptons and the Anomalous
  Magnetic Moment of the Muon}, Phys. Lett. B 116 (1982) 283--286.
\newblock \href {http://dx.doi.org/10.1016/0370-2693(82)90343-4}
  {\path{doi:10.1016/0370-2693(82)90343-4}}.

\bibitem{Grifols:1982vx}
J.~A. Grifols, A.~Mendez, {Constraints on Supersymmetric Particle Masses From
  ($g-2$) $\mu$}, Phys. Rev. D 26 (1982) 1809.
\newblock \href {http://dx.doi.org/10.1103/PhysRevD.26.1809}
  {\path{doi:10.1103/PhysRevD.26.1809}}.

\bibitem{Barbieri:1982aj}
R.~Barbieri, L.~Maiani, {The Muon Anomalous Magnetic Moment in Broken
  Supersymmetric Theories}, Phys. Lett. B 117 (1982) 203--207.
\newblock \href {http://dx.doi.org/10.1016/0370-2693(82)90547-0}
  {\path{doi:10.1016/0370-2693(82)90547-0}}.

\bibitem{Kosower:1983yw}
D.~A. Kosower, L.~M. Krauss, N.~Sakai, {Low-Energy Supergravity and the
  Anomalous Magnetic Moment of the Muon}, Phys. Lett. B 133 (1983) 305--310.
\newblock \href {http://dx.doi.org/10.1016/0370-2693(83)90152-1}
  {\path{doi:10.1016/0370-2693(83)90152-1}}.

\bibitem{Yuan:1984ww}
T.~C. Yuan, R.~L. Arnowitt, A.~H. Chamseddine, P.~Nath, {Supersymmetric
  Electroweak Effects on G-2 (mu)}, Z. Phys. C 26 (1984) 407.
\newblock \href {http://dx.doi.org/10.1007/BF01452567}
  {\path{doi:10.1007/BF01452567}}.

\bibitem{Lopez:1993vi}
J.~L. Lopez, D.~V. Nanopoulos, X.~Wang, {Large (g-2)-mu in SU(5) x U(1)
  supergravity models}, Phys. Rev. D 49 (1994) 366--372.
\newblock \href {http://arxiv.org/abs/hep-ph/9308336}
  {\path{arXiv:hep-ph/9308336}}, \href
  {http://dx.doi.org/10.1103/PhysRevD.49.366}
  {\path{doi:10.1103/PhysRevD.49.366}}.

\bibitem{Chattopadhyay:1995ae}
U.~Chattopadhyay, P.~Nath, {Probing supergravity grand unification in the
  Brookhaven g-2 experiment}, Phys. Rev. D 53 (1996) 1648--1657.
\newblock \href {http://arxiv.org/abs/hep-ph/9507386}
  {\path{arXiv:hep-ph/9507386}}, \href
  {http://dx.doi.org/10.1103/PhysRevD.53.1648}
  {\path{doi:10.1103/PhysRevD.53.1648}}.

\bibitem{Carena:1996qa}
M.~Carena, G.~F. Giudice, C.~E.~M. Wagner, {Constraints on supersymmetric
  models from the muon anomalous magnetic moment}, Phys. Lett. B 390 (1997)
  234--242.
\newblock \href {http://arxiv.org/abs/hep-ph/9610233}
  {\path{arXiv:hep-ph/9610233}}, \href
  {http://dx.doi.org/10.1016/S0370-2693(96)01396-2}
  {\path{doi:10.1016/S0370-2693(96)01396-2}}.

\bibitem{Baltz:2001ts}
E.~A. Baltz, P.~Gondolo, {Implications of muon anomalous magnetic moment for
  supersymmetric dark matter}, Phys. Rev. Lett. 86 (2001) 5004.
\newblock \href {http://arxiv.org/abs/hep-ph/0102147}
  {\path{arXiv:hep-ph/0102147}}, \href
  {http://dx.doi.org/10.1103/PhysRevLett.86.5004}
  {\path{doi:10.1103/PhysRevLett.86.5004}}.

\bibitem{Everett:2001tq}
L.~L. Everett, G.~L. Kane, S.~Rigolin, L.-T. Wang, {Implications of muon g-2
  for supersymmetry and for discovering superpartners directly}, Phys. Rev.
  Lett. 86 (2001) 3484--3487.
\newblock \href {http://arxiv.org/abs/hep-ph/0102145}
  {\path{arXiv:hep-ph/0102145}}, \href
  {http://dx.doi.org/10.1103/PhysRevLett.86.3484}
  {\path{doi:10.1103/PhysRevLett.86.3484}}.

\bibitem{Martin:2001st}
S.~P. Martin, J.~D. Wells, {Muon Anomalous Magnetic Dipole Moment in
  Supersymmetric Theories}, Phys. Rev. D 64 (2001) 035003.
\newblock \href {http://arxiv.org/abs/hep-ph/0103067}
  {\path{arXiv:hep-ph/0103067}}, \href
  {http://dx.doi.org/10.1103/PhysRevD.64.035003}
  {\path{doi:10.1103/PhysRevD.64.035003}}.

\bibitem{Baer:2001kn}
H.~Baer, C.~Balazs, J.~Ferrandis, X.~Tata, {Impact of muon anomalous magnetic
  moment on supersymmetric models}, Phys. Rev. D 64 (2001) 035004.
\newblock \href {http://arxiv.org/abs/hep-ph/0103280}
  {\path{arXiv:hep-ph/0103280}}, \href
  {http://dx.doi.org/10.1103/PhysRevD.64.035004}
  {\path{doi:10.1103/PhysRevD.64.035004}}.

\bibitem{Baer:2002gm}
H.~Baer, C.~Balazs, A.~Belyaev, J.~K. Mizukoshi, X.~Tata, Y.~Wang, {Updated
  constraints on the minimal supergravity model}, JHEP 07 (2002) 050.
\newblock \href {http://arxiv.org/abs/hep-ph/0205325}
  {\path{arXiv:hep-ph/0205325}}, \href
  {http://dx.doi.org/10.1088/1126-6708/2002/07/050}
  {\path{doi:10.1088/1126-6708/2002/07/050}}.

\bibitem{Barger:2004mr}
V.~Barger, C.~Kao, P.~Langacker, H.-S. Lee, {Muon anomalous magnetic moment in
  a supersymmetric U(1)-prime model}, Phys. Lett. B 614 (2005) 67--77.
\newblock \href {http://arxiv.org/abs/hep-ph/0412136}
  {\path{arXiv:hep-ph/0412136}}, \href
  {http://dx.doi.org/10.1016/j.physletb.2005.03.042}
  {\path{doi:10.1016/j.physletb.2005.03.042}}.

\bibitem{Czarnecki:2001pv}
A.~Czarnecki, W.~J. Marciano, {The Muon anomalous magnetic moment: A Harbinger
  for 'new physics'}, Phys. Rev. D 64 (2001) 013014.
\newblock \href {http://arxiv.org/abs/hep-ph/0102122}
  {\path{arXiv:hep-ph/0102122}}, \href
  {http://dx.doi.org/10.1103/PhysRevD.64.013014}
  {\path{doi:10.1103/PhysRevD.64.013014}}.

\bibitem{Feng:2001tr}
J.~L. Feng, K.~T. Matchev, {Supersymmetry and the anomalous magnetic moment of
  the muon}, Phys. Rev. Lett. 86 (2001) 3480--3483.
\newblock \href {http://arxiv.org/abs/hep-ph/0102146}
  {\path{arXiv:hep-ph/0102146}}, \href
  {http://dx.doi.org/10.1103/PhysRevLett.86.3480}
  {\path{doi:10.1103/PhysRevLett.86.3480}}.

\bibitem{Canepa:2019hph}
A.~Canepa, {Searches for Supersymmetry at the Large Hadron Collider}, Rev.
  Phys. 4 (2019) 100033.
\newblock \href {http://dx.doi.org/10.1016/j.revip.2019.100033}
  {\path{doi:10.1016/j.revip.2019.100033}}.

\bibitem{Aad:2019vnb}
G.~Aad, et~al., {Search for electroweak production of charginos and sleptons
  decaying into final states with two leptons and missing transverse momentum
  in $\sqrt{s}=13$ TeV $pp$ collisions using the ATLAS detector}, Eur. Phys. J.
  C 80~(2) (2020) 123.
\newblock \href {http://arxiv.org/abs/1908.08215} {\path{arXiv:1908.08215}},
  \href {http://dx.doi.org/10.1140/epjc/s10052-019-7594-6}
  {\path{doi:10.1140/epjc/s10052-019-7594-6}}.

\bibitem{Sirunyan:2020eab}
A.~M. Sirunyan, et~al., {Search for supersymmetry in final states with two
  oppositely charged same-flavor leptons and missing transverse momentum in
  proton-proton collisions at $\sqrt{s}=$ 13 TeV}\href
  {http://arxiv.org/abs/2012.08600} {\path{arXiv:2012.08600}}.

\bibitem{Baer:2011ab}
H.~Baer, V.~Barger, A.~Mustafayev, {Implications of a 125 GeV Higgs scalar for
  LHC SUSY and neutralino dark matter searches}, Phys. Rev. D 85 (2012) 075010.
\newblock \href {http://arxiv.org/abs/1112.3017} {\path{arXiv:1112.3017}},
  \href {http://dx.doi.org/10.1103/PhysRevD.85.075010}
  {\path{doi:10.1103/PhysRevD.85.075010}}.

\bibitem{Arbey:2011ab}
A.~Arbey, M.~Battaglia, A.~Djouadi, F.~Mahmoudi, J.~Quevillon, {Implications of
  a 125 GeV Higgs for supersymmetric models}, Phys. Lett. B 708 (2012)
  162--169.
\newblock \href {http://arxiv.org/abs/1112.3028} {\path{arXiv:1112.3028}},
  \href {http://dx.doi.org/10.1016/j.physletb.2012.01.053}
  {\path{doi:10.1016/j.physletb.2012.01.053}}.

\bibitem{Aad:2015pla}
G.~Aad, et~al., {Constraints on new phenomena via Higgs boson couplings and
  invisible decays with the ATLAS detector}, JHEP 11 (2015) 206.
\newblock \href {http://arxiv.org/abs/1509.00672} {\path{arXiv:1509.00672}},
  \href {http://dx.doi.org/10.1007/JHEP11(2015)206}
  {\path{doi:10.1007/JHEP11(2015)206}}.

\bibitem{Sirunyan:2018koj}
A.~M. Sirunyan, et~al., {Combined measurements of Higgs boson couplings in
  proton\textendash{}proton collisions at $\sqrt{s}=13\,\text {Te}\text {V} $},
  Eur. Phys. J. C 79~(5) (2019) 421.
\newblock \href {http://arxiv.org/abs/1809.10733} {\path{arXiv:1809.10733}},
  \href {http://dx.doi.org/10.1140/epjc/s10052-019-6909-y}
  {\path{doi:10.1140/epjc/s10052-019-6909-y}}.

\bibitem{Bae:2015nva}
K.~J. Bae, H.~Baer, N.~Nagata, H.~Serce, {Prospects for Higgs coupling
  measurements in SUSY with radiatively-driven naturalness}, Phys. Rev. D
  92~(3) (2015) 035006.
\newblock \href {http://arxiv.org/abs/1505.03541} {\path{arXiv:1505.03541}},
  \href {http://dx.doi.org/10.1103/PhysRevD.92.035006}
  {\path{doi:10.1103/PhysRevD.92.035006}}.

\bibitem{Haber:2018ltt}
H.~E. Haber, {Approximate Higgs alignment without decoupling}, in: {53rd
  Rencontres de Moriond on QCD and High Energy Interactions}, 2018.
\newblock \href {http://arxiv.org/abs/1805.05754} {\path{arXiv:1805.05754}}.

\bibitem{Bertolini:1990if}
S.~Bertolini, F.~Borzumati, A.~Masiero, G.~Ridolfi, {Effects of supergravity
  induced electroweak breaking on rare $B$ decays and mixings}, Nucl. Phys. B
  353 (1991) 591--649.
\newblock \href {http://dx.doi.org/10.1016/0550-3213(91)90320-W}
  {\path{doi:10.1016/0550-3213(91)90320-W}}.

\bibitem{Misiak:2015xwa}
M.~Misiak, et~al., {Updated NNLO QCD predictions for the weak radiative B-meson
  decays}, Phys. Rev. Lett. 114~(22) (2015) 221801.
\newblock \href {http://arxiv.org/abs/1503.01789} {\path{arXiv:1503.01789}},
  \href {http://dx.doi.org/10.1103/PhysRevLett.114.221801}
  {\path{doi:10.1103/PhysRevLett.114.221801}}.

\bibitem{Belle:2016ufb}
A.~Abdesselam, et~al., {Measurement of the inclusive $B\to X_{s+d} \gamma$
  branching fraction, photon energy spectrum and HQE parameters}, in: {38th
  International Conference on High Energy Physics}, 2016.
\newblock \href {http://arxiv.org/abs/1608.02344} {\path{arXiv:1608.02344}}.

\bibitem{Baer:2016bwh}
H.~Baer, V.~Barger, N.~Nagata, M.~Savoy, {Phenomenological profile of top
  squarks from natural supersymmetry at the LHC}, Phys. Rev. D 95~(5) (2017)
  055012, [Addendum: Phys.Rev.D 103, 059902 (2021)].
\newblock \href {http://arxiv.org/abs/1611.08511} {\path{arXiv:1611.08511}},
  \href {http://dx.doi.org/10.1103/PhysRevD.95.055012}
  {\path{doi:10.1103/PhysRevD.95.055012}}.

\bibitem{Gabbiani:1996hi}
F.~Gabbiani, E.~Gabrielli, A.~Masiero, L.~Silvestrini, {A Complete analysis of
  FCNC and CP constraints in general SUSY extensions of the standard model},
  Nucl. Phys. B 477 (1996) 321--352.
\newblock \href {http://arxiv.org/abs/hep-ph/9604387}
  {\path{arXiv:hep-ph/9604387}}, \href
  {http://dx.doi.org/10.1016/0550-3213(96)00390-2}
  {\path{doi:10.1016/0550-3213(96)00390-2}}.

\bibitem{Nir:1993mx}
Y.~Nir, N.~Seiberg, {Should squarks be degenerate?}, Phys. Lett. B 309 (1993)
  337--343.
\newblock \href {http://arxiv.org/abs/hep-ph/9304307}
  {\path{arXiv:hep-ph/9304307}}, \href
  {http://dx.doi.org/10.1016/0370-2693(93)90942-B}
  {\path{doi:10.1016/0370-2693(93)90942-B}}.

\bibitem{Dine:1993np}
M.~Dine, R.~G. Leigh, A.~Kagan, {Flavor symmetries and the problem of squark
  degeneracy}, Phys. Rev. D 48 (1993) 4269--4274.
\newblock \href {http://arxiv.org/abs/hep-ph/9304299}
  {\path{arXiv:hep-ph/9304299}}, \href
  {http://dx.doi.org/10.1103/PhysRevD.48.4269}
  {\path{doi:10.1103/PhysRevD.48.4269}}.

\bibitem{Cohen:1996vb}
A.~G. Cohen, D.~B. Kaplan, A.~E. Nelson, {The More minimal supersymmetric
  standard model}, Phys. Lett. B 388 (1996) 588--598.
\newblock \href {http://arxiv.org/abs/hep-ph/9607394}
  {\path{arXiv:hep-ph/9607394}}, \href
  {http://dx.doi.org/10.1016/S0370-2693(96)01183-5}
  {\path{doi:10.1016/S0370-2693(96)01183-5}}.

\bibitem{Baer:2014ica}
H.~Baer, V.~Barger, D.~Mickelson, M.~Padeffke-Kirkland, {SUSY models under
  siege: LHC constraints and electroweak fine-tuning}, Phys. Rev. D 89~(11)
  (2014) 115019.
\newblock \href {http://arxiv.org/abs/1404.2277} {\path{arXiv:1404.2277}},
  \href {http://dx.doi.org/10.1103/PhysRevD.89.115019}
  {\path{doi:10.1103/PhysRevD.89.115019}}.

\bibitem{Baer:2018hwa}
H.~Baer, V.~Barger, D.~Sengupta, {Anomaly mediated SUSY breaking model
  retrofitted for naturalness}, Phys. Rev. D 98~(1) (2018) 015039.
\newblock \href {http://arxiv.org/abs/1801.09730} {\path{arXiv:1801.09730}},
  \href {http://dx.doi.org/10.1103/PhysRevD.98.015039}
  {\path{doi:10.1103/PhysRevD.98.015039}}.

\bibitem{Khlopov:1984pf}
M.~Y. Khlopov, A.~D. Linde, {Is It Easy to Save the Gravitino?}, Phys. Lett. B
  138 (1984) 265--268.
\newblock \href {http://dx.doi.org/10.1016/0370-2693(84)91656-3}
  {\path{doi:10.1016/0370-2693(84)91656-3}}.

\bibitem{Kawasaki:2008qe}
M.~Kawasaki, K.~Kohri, T.~Moroi, A.~Yotsuyanagi, {Big-Bang Nucleosynthesis and
  Gravitino}, Phys. Rev. D 78 (2008) 065011.
\newblock \href {http://arxiv.org/abs/0804.3745} {\path{arXiv:0804.3745}},
  \href {http://dx.doi.org/10.1103/PhysRevD.78.065011}
  {\path{doi:10.1103/PhysRevD.78.065011}}.

\bibitem{Linde:1996cx}
A.~D. Linde, {Relaxing the cosmological moduli problem}, Phys. Rev. D 53 (1996)
  R4129--R4132.
\newblock \href {http://arxiv.org/abs/hep-th/9601083}
  {\path{arXiv:hep-th/9601083}}, \href
  {http://dx.doi.org/10.1103/PhysRevD.53.R4129}
  {\path{doi:10.1103/PhysRevD.53.R4129}}.

\bibitem{Baer:2012up}
H.~Baer, V.~Barger, P.~Huang, A.~Mustafayev, X.~Tata, {Radiative natural SUSY
  with a 125 GeV Higgs boson}, Phys. Rev. Lett. 109 (2012) 161802.
\newblock \href {http://arxiv.org/abs/1207.3343} {\path{arXiv:1207.3343}},
  \href {http://dx.doi.org/10.1103/PhysRevLett.109.161802}
  {\path{doi:10.1103/PhysRevLett.109.161802}}.

\bibitem{Baer:2012cf}
H.~Baer, V.~Barger, P.~Huang, D.~Mickelson, A.~Mustafayev, X.~Tata, {Radiative
  natural supersymmetry: Reconciling electroweak fine-tuning and the Higgs
  boson mass}, Phys. Rev. D 87~(11) (2013) 115028.
\newblock \href {http://arxiv.org/abs/1212.2655} {\path{arXiv:1212.2655}},
  \href {http://dx.doi.org/10.1103/PhysRevD.87.115028}
  {\path{doi:10.1103/PhysRevD.87.115028}}.

\bibitem{Douglas:2006es}
M.~R. Douglas, S.~Kachru, {Flux compactification}, Rev. Mod. Phys. 79 (2007)
  733--796.
\newblock \href {http://arxiv.org/abs/hep-th/0610102}
  {\path{arXiv:hep-th/0610102}}, \href
  {http://dx.doi.org/10.1103/RevModPhys.79.733}
  {\path{doi:10.1103/RevModPhys.79.733}}.

\bibitem{Agrawal:1998xa}
V.~Agrawal, S.~M. Barr, J.~F. Donoghue, D.~Seckel, {Anthropic considerations in
  multiple domain theories and the scale of electroweak symmetry breaking},
  Phys. Rev. Lett. 80 (1998) 1822--1825.
\newblock \href {http://arxiv.org/abs/hep-ph/9801253}
  {\path{arXiv:hep-ph/9801253}}, \href
  {http://dx.doi.org/10.1103/PhysRevLett.80.1822}
  {\path{doi:10.1103/PhysRevLett.80.1822}}.

\bibitem{Baer:2020kwz}
H.~Baer, V.~Barger, S.~Salam, D.~Sengupta, K.~Sinha, {Status of weak scale
  supersymmetry after LHC Run 2 and ton-scale noble liquid WIMP searches}, Eur.
  Phys. J. ST 229~(21) (2020) 3085--3141.
\newblock \href {http://arxiv.org/abs/2002.03013} {\path{arXiv:2002.03013}},
  \href {http://dx.doi.org/10.1140/epjst/e2020-000020-x}
  {\path{doi:10.1140/epjst/e2020-000020-x}}.

\bibitem{Baer:2004xx}
H.~Baer, A.~Belyaev, T.~Krupovnickas, A.~Mustafayev, {SUSY normal scalar mass
  hierarchy reconciles (g-2)(mu), b ---\ensuremath{>} s gamma and relic
  density}, JHEP 06 (2004) 044.
\newblock \href {http://arxiv.org/abs/hep-ph/0403214}
  {\path{arXiv:hep-ph/0403214}}, \href
  {http://dx.doi.org/10.1088/1126-6708/2004/06/044}
  {\path{doi:10.1088/1126-6708/2004/06/044}}.

\bibitem{Yamaguchi:2016oqz}
M.~Yamaguchi, W.~Yin, {A novel approach to finely tuned supersymmetric standard
  models: The case of the non-universal Higgs mass model}, PTEP 2018~(2) (2018)
  023B06.
\newblock \href {http://arxiv.org/abs/1606.04953} {\path{arXiv:1606.04953}},
  \href {http://dx.doi.org/10.1093/ptep/pty002}
  {\path{doi:10.1093/ptep/pty002}}.

\bibitem{nuhm2}
D.~Matalliotakis, H.~Nilles,
  \href{http://dx.doi.org/10.1016/0550-3213(94)00487-Y}{Implications of
  non-universality of soft terms in supersymmetric grand unified theories},
  Nuclear Physics B 435~(1-2) (1995) 115–128.
\newblock \href {http://dx.doi.org/10.1016/0550-3213(94)00487-y}
  {\path{doi:10.1016/0550-3213(94)00487-y}}.
\newline\urlprefix\url{http://dx.doi.org/10.1016/0550-3213(94)00487-Y}

\bibitem{nuhm22}
M.~Olechowski, S.~Pokorski,
  \href{http://dx.doi.org/10.1016/0370-2693(94)01571-S}{Electroweak symmetry
  breaking with non-universal scalar soft terms and large tan β solutions},
  Physics Letters B 344~(1-4) (1995) 201–210.
\newblock \href {http://dx.doi.org/10.1016/0370-2693(94)01571-s}
  {\path{doi:10.1016/0370-2693(94)01571-s}}.
\newline\urlprefix\url{http://dx.doi.org/10.1016/0370-2693(94)01571-S}

\bibitem{nuhm23}
P.~Nath, R.~Arnowitt,
  \href{http://dx.doi.org/10.1142/9789814447263-0020}{Non-universal soft susy
  breaking and dark matter}, COSMO-97\href
  {http://dx.doi.org/10.1142/9789814447263-0020}
  {\path{doi:10.1142/9789814447263-0020}}.
\newline\urlprefix\url{http://dx.doi.org/10.1142/9789814447263-0020}

\bibitem{nuhm24}
J.~Ellis, K.~Olive, Y.~Santoso,
  \href{http://dx.doi.org/10.1016/S0370-2693(02)02071-3}{The mssm parameter
  space with non-universal higgs masses}, Physics Letters B 539~(1-2) (2002)
  107–118.
\newblock \href {http://dx.doi.org/10.1016/s0370-2693(02)02071-3}
  {\path{doi:10.1016/s0370-2693(02)02071-3}}.
\newline\urlprefix\url{http://dx.doi.org/10.1016/S0370-2693(02)02071-3}

\bibitem{nuhm25}
J.~Ellis, T.~Falk, K.~A. Olive, Y.~Santoso,
  \href{http://dx.doi.org/10.1016/S0550-3213(02)01144-6}{Exploration of the
  mssm with non-universal higgs masses}, Nuclear Physics B 652 (2003)
  259–347.
\newblock \href {http://dx.doi.org/10.1016/s0550-3213(02)01144-6}
  {\path{doi:10.1016/s0550-3213(02)01144-6}}.
\newline\urlprefix\url{http://dx.doi.org/10.1016/S0550-3213(02)01144-6}

\bibitem{nuhm26}
H.~Baer, A.~Mustafayev, S.~Profumo, A.~Belyaev, X.~Tata, {Direct, indirect and
  collider detection of neutralino dark matter in SUSY models with
  non-universal Higgs masses}, JHEP 07 (2005) 065.
\newblock \href {http://arxiv.org/abs/hep-ph/0504001}
  {\path{arXiv:hep-ph/0504001}}, \href
  {http://dx.doi.org/10.1088/1126-6708/2005/07/065}
  {\path{doi:10.1088/1126-6708/2005/07/065}}.

\bibitem{isajet}
F.~E. Paige, S.~D. Protopopescu, H.~Baer, X.~Tata, {ISAJET 7.69: A Monte Carlo
  event generator for pp, anti-p p, and e+e- reactions}\href
  {http://arxiv.org/abs/hep-ph/0312045} {\path{arXiv:hep-ph/0312045}}.

\bibitem{Endo:2021zal}
M.~Endo, K.~Hamaguchi, S.~Iwamoto, T.~Kitahara, {Supersymmetric Interpretation
  of the Muon $g-2$ Anomaly}\href {http://arxiv.org/abs/2104.03217}
  {\path{arXiv:2104.03217}}.

\bibitem{Iwamoto:2021aaf}
S.~Iwamoto, T.~T. Yanagida, N.~Yokozaki, {Wino-Higgsino dark matter in the MSSM
  from the $g-2$ anomaly}\href {http://arxiv.org/abs/2104.03223}
  {\path{arXiv:2104.03223}}.

\bibitem{Gu:2021mjd}
Y.~Gu, N.~Liu, L.~Su, D.~Wang, {Heavy Bino and Slepton for Muon g-2
  Anomaly}\href {http://arxiv.org/abs/2104.03239} {\path{arXiv:2104.03239}}.

\bibitem{VanBeekveld:2021tgn}
M.~Van~Beekveld, W.~Beenakker, M.~Schutten, J.~De~Wit, {Dark matter,
  fine-tuning and $(g-2)_{\mu}$ in the pMSSM}\href
  {http://arxiv.org/abs/2104.03245} {\path{arXiv:2104.03245}}.

\bibitem{Yin:2021mls}
W.~Yin, {Muon $g-2$ Anomaly in Anomaly Mediation}\href
  {http://arxiv.org/abs/2104.03259} {\path{arXiv:2104.03259}}.

\bibitem{Wang:2021bcx}
F.~Wang, L.~Wu, Y.~Xiao, J.~M. Yang, Y.~Zhang, {GUT-scale constrained SUSY in
  light of E989 muon g-2 measurement}\href {http://arxiv.org/abs/2104.03262}
  {\path{arXiv:2104.03262}}.

\bibitem{Abdughani:2021pdc}
M.~Abdughani, Y.-Z. Fan, L.~Feng, Y.-L. Sming~Tsai, L.~Wu, Q.~Yuan, {A common
  origin of muon g-2 anomaly, Galaxy Center GeV excess and AMS-02 anti-proton
  excess in the NMSSM}\href {http://arxiv.org/abs/2104.03274}
  {\path{arXiv:2104.03274}}.

\bibitem{Cao:2021tuh}
J.~Cao, J.~Lian, Y.~Pan, D.~Zhang, P.~Zhu, {Imporved $(g-2)_\mu$ Measurement
  and Singlino dark matter in the general NMSSM}\href
  {http://arxiv.org/abs/2104.03284} {\path{arXiv:2104.03284}}.

\bibitem{Chakraborti:2021dli}
M.~Chakraborti, S.~Heinemeyer, I.~Saha, {The new ''MUON G-2'' Result and
  Supersymmetry}\href {http://arxiv.org/abs/2104.03287}
  {\path{arXiv:2104.03287}}.

\bibitem{Ibe:2021cvf}
M.~Ibe, S.~Kobayashi, Y.~Nakayama, S.~Shirai, {Muon $g-2$ in Gauge Mediation
  without SUSY CP Problem}\href {http://arxiv.org/abs/2104.03289}
  {\path{arXiv:2104.03289}}.

\bibitem{Cox:2021gqq}
P.~Cox, C.~Han, T.~T. Yanagida, {Muon $g-2$ and Co-annihilating Dark Matter in
  the MSSM}\href {http://arxiv.org/abs/2104.03290} {\path{arXiv:2104.03290}}.

\bibitem{Han:2021ify}
C.~Han, {Muon g-2 and CP violation in MSSM}\href
  {http://arxiv.org/abs/2104.03292} {\path{arXiv:2104.03292}}.

\bibitem{Heinemeyer:2021zpc}
S.~Heinemeyer, E.~Kpatcha, I.~n. Lara, D.~E. L\'opez-Fogliani, C.~Mu\~noz,
  N.~Nagata, {The new $(g-2)_\mu$ result and the $\mu\nu$SSM}\href
  {http://arxiv.org/abs/2104.03294} {\path{arXiv:2104.03294}}.

\bibitem{Baum:2021qzx}
S.~Baum, M.~Carena, N.~R. Shah, C.~E.~M. Wagner, {The Tiny (g-2) Muon Wobble
  from Small-$\mu$ Supersymmetry}\href {http://arxiv.org/abs/2104.03302}
  {\path{arXiv:2104.03302}}.

\bibitem{Zhang:2021gun}
H.-B. Zhang, C.-X. Liu, J.-L. Yang, T.-F. Feng, {Muon anomalous magnetic dipole
  moment in the $\mu\nu$SSM}\href {http://arxiv.org/abs/2104.03489}
  {\path{arXiv:2104.03489}}.

\bibitem{Ahmed:2021htr}
W.~Ahmed, I.~Khan, J.~Li, T.~Li, S.~Raza, W.~Zhang, {The Natural Explanation of
  the Muon Anomalous Magnetic Moment via the Electroweak Supersymmetry from the
  GmSUGRA in the MSSM}\href {http://arxiv.org/abs/2104.03491}
  {\path{arXiv:2104.03491}}.

\bibitem{Athron:2021iuf}
P.~Athron, C.~Bal\'azs, D.~H. Jacob, W.~Kotlarski, D.~St\"ockinger,
  H.~St\"ockinger-Kim, {New physics explanations of $a_\mu$ in light of the
  FNAL muon $g-2$ measurement}\href {http://arxiv.org/abs/2104.03691}
  {\path{arXiv:2104.03691}}.

\bibitem{Aboubrahim:2021rwz}
A.~Aboubrahim, M.~Klasen, P.~Nath, {What Fermilab $(g-2)_{\mu}$ experiment
  tells us about discovering SUSY at HL-LHC and HE-LHC}\href
  {http://arxiv.org/abs/2104.03839} {\path{arXiv:2104.03839}}.

\bibitem{Chakraborti:2021bmv}
M.~Chakraborti, L.~Roszkowski, S.~Trojanowski, {GUT-constrained supersymmetry
  and dark matter in light of the new $(g-2)_\mu$ determination}\href
  {http://arxiv.org/abs/2104.04458} {\path{arXiv:2104.04458}}.

\bibitem{pbmz}
D.~M. Pierce, J.~A. Bagger, K.~T. Matchev, R.-j. Zhang, {Precision corrections
  in the minimal supersymmetric standard model}, Nucl. Phys. B 491 (1997)
  3--67.
\newblock \href {http://arxiv.org/abs/hep-ph/9606211}
  {\path{arXiv:hep-ph/9606211}}, \href
  {http://dx.doi.org/10.1016/S0550-3213(96)00683-9}
  {\path{doi:10.1016/S0550-3213(96)00683-9}}.

\bibitem{Baer:2011hx}
H.~Baer, A.~Lessa, S.~Rajagopalan, W.~Sreethawong, {Mixed axion/neutralino cold
  dark matter in supersymmetric models}, JCAP 06 (2011) 031.
\newblock \href {http://arxiv.org/abs/1103.5413} {\path{arXiv:1103.5413}},
  \href {http://dx.doi.org/10.1088/1475-7516/2011/06/031}
  {\path{doi:10.1088/1475-7516/2011/06/031}}.

\bibitem{Lindner:2016bgg}
M.~Lindner, M.~Platscher, F.~S. Queiroz, {A Call for New Physics : The Muon
  Anomalous Magnetic Moment and Lepton Flavor Violation}, Phys. Rept. 731
  (2018) 1--82.
\newblock \href {http://arxiv.org/abs/1610.06587} {\path{arXiv:1610.06587}},
  \href {http://dx.doi.org/10.1016/j.physrep.2017.12.001}
  {\path{doi:10.1016/j.physrep.2017.12.001}}.

\bibitem{Baer:2019xug}
H.~Baer, V.~Barger, H.~Serce, {Lepton flavor violation from SUSY with
  non-universal scalars}, Phys. Rev. Research. 1 (2019) 033022.
\newblock \href {http://arxiv.org/abs/1907.06693} {\path{arXiv:1907.06693}},
  \href {http://dx.doi.org/10.1103/PhysRevResearch.1.033022}
  {\path{doi:10.1103/PhysRevResearch.1.033022}}.

\bibitem{CidVidal:2018eel}
X.~Cid~Vidal, et~al., {Report from Working Group 3}: {Beyond the Standard Model
  physics at the HL-LHC and HE-LHC}, CERN Yellow Rep. Monogr. 7 (2019)
  585--865.
\newblock \href {http://arxiv.org/abs/1812.07831} {\path{arXiv:1812.07831}},
  \href {http://dx.doi.org/10.23731/CYRM-2019-007.585}
  {\path{doi:10.23731/CYRM-2019-007.585}}.

\bibitem{Susskind:2003kw}
L.~Susskind, {The Anthropic landscape of string theory}\href
  {http://arxiv.org/abs/hep-th/0302219} {\path{arXiv:hep-th/0302219}}.

\bibitem{Weinberg:1987dv}
S.~Weinberg, {Anthropic Bound on the Cosmological Constant}, Phys. Rev. Lett.
  59 (1987) 2607.
\newblock \href {http://dx.doi.org/10.1103/PhysRevLett.59.2607}
  {\path{doi:10.1103/PhysRevLett.59.2607}}.

\bibitem{Douglas:2004qg}
M.~R. Douglas, {Statistical analysis of the supersymmetry breaking scale}\href
  {http://arxiv.org/abs/hep-th/0405279} {\path{arXiv:hep-th/0405279}}.

\bibitem{Baer:2017uvn}
H.~Baer, V.~Barger, H.~Serce, K.~Sinha, {Higgs and superparticle mass
  predictions from the landscape}, JHEP 03 (2018) 002.
\newblock \href {http://arxiv.org/abs/1712.01399} {\path{arXiv:1712.01399}},
  \href {http://dx.doi.org/10.1007/JHEP03(2018)002}
  {\path{doi:10.1007/JHEP03(2018)002}}.

\bibitem{Baer:2020dri}
H.~Baer, V.~Barger, S.~Salam, D.~Sengupta, {Landscape Higgs boson and sparticle
  mass predictions from a logarithmic soft term distribution}, Phys. Rev. D
  103~(3) (2021) 035031.
\newblock \href {http://arxiv.org/abs/2011.04035} {\path{arXiv:2011.04035}},
  \href {http://dx.doi.org/10.1103/PhysRevD.103.035031}
  {\path{doi:10.1103/PhysRevD.103.035031}}.

\bibitem{Baer:2020vad}
H.~Baer, V.~Barger, S.~Salam, D.~Sengupta, {String landscape guide to soft SUSY
  breaking terms}, Phys. Rev. D 102~(7) (2020) 075012.
\newblock \href {http://arxiv.org/abs/2005.13577} {\path{arXiv:2005.13577}},
  \href {http://dx.doi.org/10.1103/PhysRevD.102.075012}
  {\path{doi:10.1103/PhysRevD.102.075012}}.

\bibitem{Baer:2019zfl}
H.~Baer, V.~Barger, D.~Sengupta, {Landscape solution to the SUSY flavor and CP
  problems}, Phys. Rev. Res. 1~(3) (2019) 033179.
\newblock \href {http://arxiv.org/abs/1910.00090} {\path{arXiv:1910.00090}},
  \href {http://dx.doi.org/10.1103/PhysRevResearch.1.033179}
  {\path{doi:10.1103/PhysRevResearch.1.033179}}.

\bibitem{Hagiwara:2017lse}
K.~Hagiwara, K.~Ma, S.~Mukhopadhyay, {Closing in on the chargino contribution
  to the muon g-2 in the MSSM: current LHC constraints}, Phys. Rev. D 97~(5)
  (2018) 055035.
\newblock \href {http://arxiv.org/abs/1706.09313} {\path{arXiv:1706.09313}},
  \href {http://dx.doi.org/10.1103/PhysRevD.97.055035}
  {\path{doi:10.1103/PhysRevD.97.055035}}.

\end{thebibliography}
\bibliographystyle{elsarticle-num}

\end{document}